\documentclass[preprint,prl,showpacs,amsmath,%
nofootinbib,amssymb,showkeys,byrevtex,aps]{revtex4}
\usepackage{graphicx}
\usepackage{amsmath}
\usepackage{bm}
\newcommand{\be}{\begin{equation}}
\newcommand{\ee}{\end{equation}}

\begin{document}

%\preprint{}

\title{Electromagnetic shape resonances of a dielectric sphere
and radiation of portable telephones}


\author{V.~V.~Nesterenko}
\affiliation{Bogoliubov Laboratory of Theoretical Physics, Joint
Institute for Nuclear Research, 141980 Dubna, Russia}
\email{nestr@theor.jinr.ru}
\author{A.~Feoli}
\affiliation{Dipartimento di Ingegneria, Universit\`{a} del
Sannio, Corso Garibaldi n.\ 107, Palazzo Bosco Lucarelli,  82100
Benevento,  Italy} \affiliation{INFN Sezione di Napoli, Gruppo
collegato di Salerno 80126 Napoli, Italy}
\email{feoli@unisannio.it}
 % ; fax: +39089965275}}
\author{G.~Lambiase}
%\affiliation{Dipartimento di Fisica "E.R.Caianiello" --
%Universit\`a di Salerno,
%  84081 Baronissi (SA), Italy}
%\affiliation{INFN Sezione di Napoli, Gruppo collegato di Salerno
%80126 Napoli, Italy}
\email{lambiase@sa.infn.it}
\author{G.~Scarpetta}
\affiliation{Dipartimento di Fisica "E.R.Caianiello" --
Universit\`a di Salerno,
  84081 Baronissi (SA), Italy
and \\ INFN Sezione di Napoli, Gruppo collegato di Salerno
80126 Napoli, Italy}
\email{scarpetta@sa.infn.it}

\date{\today}
%\thanks{}


\date{\today}

\begin{abstract}
The frequency band used by cellular telephones includes the
eigenfrequencies of a dielectric sphere with physical
characteristics close to those of a human head. Proceeding from
the spatial features of the natural modes of such a sphere we
propose an independent and clear evident accuracy test for the
complicated numerical calculations which are conducted when
estimating the potential hazard due to the use of cellular
telephones, in particular, for the check of a proper handling of
the electromagnetic shape resonances  of a human head.
\end{abstract}

%\pacs{41.20.Jb Electromagnetic wave propagation; radiowave propagation;\\
% \phantom{ pacs numberss }42.60.Da
% Resonators, cavities, amplifiers, arrays, and rings}

\maketitle
%\section{Introduction}
%\label{Intr}
{\bf Introduction.} Estimation of the potential health hazard
due to the use of cellular telephones is now a problem of
primary importance in view of extremely rapid development and
very wide spread of this communication aid. The safety
guidelines in this field~\cite{Health} are based on the findings
from animal experiments that the biological hazards due to radio
waves result mainly from the temperature rise in
tissues\footnote{In principle, non-ionizing radiation can lead
also to other effects in biological tissues~\cite{Sernelius}.}
and a whole-body-averaged specific absorption rate (SAR) below
0.4~W/kg is not hazardous to human health. This corresponds to a
limits on the output from the cellular phones (0.6~W at 900~MHz
frequency band and 0.27~W at 1.5~GHz frequency band). Obviously,
the {\it local} absorption rate should be also considered
especially in a human head~\cite{WF}. Theoretical estimation of
the temperature rise in tissues of a human body are accomplished
in the following way.  First the electric and magnetic fields
inside the human body are calculated by solving the Maxwell
equations with a given source (antenna of a portable telephone).
The electric field gives rise to conduction currents with the
energy dissipation rate $\sigma \,E^2/2$, where $\sigma $ is the
conduction constant of respective tissues. In turn it leads to
the temperature increase. The second step is the solution of the
respective heat conduction equation (or more precisely, bioheat
equation~\cite{WF}) with local heat sources $\sigma \,E^2/2$ and
with allowance for all the possible heat currents. Hence, for
this method the distribution of electric field inside the human
body (especially inside the head) is of primary importance.

In this letter we are going to argue that the knowledge of the
properties of electromagnetic modes for a dielectric sphere with
physical characteristics close to those for a human head can be
useful, for example, when developing an independent accuracy
test of complicated numerical calculations mentioned above. The
point is the eigenfrequencies of such a dielectric sphere lie in
the GSM 400 MHz frequency band which has been used in a first
generation of mobile phone systems and now is considered for
further use. Obviously, the natural modes of a human head belong
to this band too. The natural modes of a dielectric sphere can
be divided into two types, surface and volume modes. For the
volume modes the electromagnetic energy is distributed in the
whole volume of the sphere while in the case of surface modes
this energy is located close by the sphere surface. All this
holds for the natural modes of a human head also, however we
have no respective analytic formulas in this case.

In order to be fully confident, that the pertinent numerical
schemes handle the resonances of a human head in a proper way,
we propose an independent accuracy test  of these calculations.
Without such a check it is not obvious because the routines of
numerical solving the partial differential equations are local
ones while the spatial behaviour of the relevant eigenfunctions
characterizes the system as a whole.

{\bf Shape resonances of a dielectric sphere.}
%\label{Shape}
Let us consider a sphere of radius $a$, consisting of a material
which is characterized by permittivity $\varepsilon_1 $ and
permeability $\mu_1$. The sphere is  placed in an infinite medium
with permittivity $\varepsilon_2 $ and permeability $\mu_2$. It
is assumed also that the electric currents are absent in both the
media. The finite conductivity of the material inside  a sphere
will be taken into account below.

It is known that in the source-free case $({\bf j}=0,\;\rho =0)$
the general solution of Maxwell's equations are obtained from
two scalar functions which may be chosen in different
ways~\cite{Whittaker,Nisbet}. In the case of spherical symmetry
these functions are the scalar Debye potentials $\psi$ (see, for
example, the textbooks \cite{Stratton,Jackson}):
\begin{eqnarray}
 \mathbf{E}^{{\rm TM}}_{lm}&=&\bm{\nabla} \times
\bm{\nabla}\times(\mathbf{r}\psi^{{\rm TM}}_{lm}),\quad
\mathbf{H}^{{\rm TM}}_{lm}=-i\,\omega \,\bm{\nabla} \times
(\mathbf{r}\psi^{{\rm TM}}_{lm})\quad ({\rm TM-modes}), \nonumber \\
 \mathbf{E}^{{\rm TE}}_{lm}&=& i\,\omega \,\bm{\nabla} \times
(\mathbf{r}\psi^{{\rm TE}}_{lm}),\quad \mathbf{H}^{{\rm
TE}}_{lm}=\bm{\nabla} \times \bm{\nabla}
\times(\mathbf{r}\psi^{{\rm TE}}_{lm})\quad ({\rm
TE-modes})\,{.} \label{3-1}
\end{eqnarray}
The time dependence factor $e^{-i \omega t}$ is dropped for simplicity.
These potentials obey the Helmholtz equation inside  and outside the sphere
$(r\neq a)$ and have the
indicated angular dependence
\begin{equation}
\label{3-2} ( \bm{\nabla} ^2+k^2_i )\psi_{lm}=0,\quad
k_i^2=\varepsilon_i\,\mu_i\,\frac{\omega^2}{c^2}, \quad i=1,2,
\quad
 \psi_{lm}(\mathbf{r})=f_l(r)Y_{lm}(\Omega){.}
\end{equation}

Equations  (\ref{3-2}) should be supplemented by the boundary
conditions at the origin, at the sphere surface and at infinity.
In order for the fields to be finite at $r=0$ the Debye
potentials should be regular there. Our goal is to find the
eigenfrequencies and eigenfunctions in the problem at hand.
Therefore at the spatial infinity,   radiation conditions should
be imposed~\cite{Sommerfeld,NFLS}
\begin{equation}
\label{2-5} \lim_{r\to \infty} r^{{1}/{2}}f_{l}(r)=
\mbox{const}\,{,}\qquad  \lim_{r\to\infty}r^{{1}/{2}}\left
( \frac{\partial f_{l}(r)}{\partial r} -i k_2 f_{l}(r)\right )= 0\,{.}
\end{equation}
At the sphere surface
the standard matching conditions for electric and magnetic fields
should be satisfied~\cite{Stratton}.

In view of all this the Helmholtz equation (\ref{3-2}) becomes now
the spectral problem for the Laplace operator
multiplied by the discontinuous factor $-1/(\varepsilon_i
\,\mu_i)$
\begin{equation}
\label{3-2a}-\, \frac{1}{\varepsilon_i\,\mu_i}\,\Delta
\,\psi_{\omega l m}(r)=\frac{\omega^2}{c^2}\,\psi_{\omega l m}(r),
\quad r\neq a, \quad i=1,2\,    {.}
\end{equation}
In this problem the spectral parameter is $\omega^2/c^2$. Due to
the radiation conditions (\ref{2-5}) this parameter is
complex~\cite{NFLS,Bar}. Thus we are dealing here with shape
resonances of a dielectric sphere and the respective
eigenfunctions are the quasi-normal
modes~\cite{Nollert,FN,KS,NFLS}.

In order to obey the boundary conditions at the origin and at
spatial infinity formulated above, the solution to the spectral
problem (\ref{3-2a}) should have the form
\begin{equation}
\label{3-2b} f_{\omega l}(r)=C_1\,j_l(k_1r)\,{,} \quad r<a,\quad
f_{\omega l}(r)=C_2\,h^{(1)}_l(k_2r)\,{,} \quad r>a\,{,}
\end{equation}
where $j_l(z)$ is the spherical Bessel function and $h_l^{(1)}(z)$
is the spherical Hankel function of the first kind \cite{AS}, the
latter obeys the radiation conditions (\ref{2-5}).

At the sphere surface the tangential components of electric and
magnetic fields (\ref{3-1}) are continuous.
As a result, the eigenfrequencies
of electromagnetic field for this configuration are determined
\cite{Stratton} by the frequency equation for the TE-modes
\begin{equation}
\label{3-3} \sqrt{\varepsilon_1\mu_2}\,\hat j_l'(k_1a)\,\hat
h_l(k_2a)- \sqrt{\varepsilon_2\mu_1}\,\hat j_l(k_1a)\,{\hat
h_l}'(k_2a)=0\,{}
\end{equation}
and by the analogous equation for the TM-modes
\begin{equation}
\label{3-4} \sqrt{\varepsilon_2\mu_1}\,\hat j_l'(k_1a)\,\hat
h_l(k_2a)- \sqrt{\varepsilon_1\mu_2}\,\hat j_l(k_1a)\,{\hat
h_l}'(k_2a)=0\,{,}
\end{equation}
where $k_i=\sqrt{\varepsilon_i\mu_i}\,\omega/c,\quad i=1,2$ are
the wave numbers inside and outside the sphere, respectively, and
$\hat j_l(z)$ and $\hat h_l(z)$ are the Riccati-Bessel
functions~\cite{AS}
\begin{equation}
\label{3-5} \hat j_l(z)=z\,j_l(z)\,{,}\quad \hat
h_l(z)=z\,h_l^{(1)}(z)\,{.}
\end{equation}
In equations (\ref{3-3}) and (\ref{3-4}) the orbital momentum $l$
assumes the values $1,2,\ldots $, and prime stands for the
differentiation with respect of the arguments $k_1a$ and $k_2a$ of
the Riccati-Bessel functions.

The frequency equations for a dielectric sphere of permittivity
$\varepsilon $ placed in vacuum follow from (\ref{3-3}) and
(\ref{3-4}) after putting there
\begin{equation}
\label{3-12} \varepsilon_1=\varepsilon, \quad
\varepsilon_2=\mu_1=\mu_2=1\,{.}
\end{equation}
The roots of these equations have been studied in the Debye paper
\cite{Debye} by making use of an approximate  method. As the
starting solution the eigenfrequencies of a perfectly conducting
sphere were used. In this case the frequencies are different for
electromagnetic oscillations inside and outside  sphere. Namely,
inside  sphere they are given by the roots of the
equations $(l\geq 1)$
\begin{eqnarray}
j_l\left ( \frac{\omega}{c}\,a \right)&=&0 \quad
(\mbox{TE-modes})\,{,} \label{3-13}\\ \frac{d}{dr}
\left(r\,j_l\left ( \frac{\omega}{c}\,a \right)\right)&=&0\,{,}
\quad r=a \quad (\mbox{TM-modes})\,{,} \label{3-14}
\end{eqnarray}
while outside  sphere these frequencies  are determined by equations
\begin{eqnarray}
\label{2-7} h^{(1)}_l\left ( \frac{\omega}{c}\,a \right
)&=&0 \quad (\mbox{TE-modes}) \\
 \frac{d}{d r}\left ( r \,h^{(1)}_l\left (
\frac{\omega}{c}\,r \right ) \right )&=&0, \quad r=a \quad (\mbox{TM-modes})\,{.}
\label{2-8}
\end{eqnarray}
The frequency equations for perfectly conducting
sphere  (\ref{3-13}), (\ref{3-14}) and (\ref{2-7}), (\ref{2-8})
can be formally derived  by substituting (\ref{3-12}) into
frequency equations (\ref{3-3}) and (\ref{3-4}) and taking there
the limit $\varepsilon \to \infty$.

 Approximate calculation of the eigenfrequencies of a dielectric
sphere without using computer \cite{Debye} didnot allow one to
reveal the characteristic features of the respective
eigenfunctions (quasi-normal modes). The computer analysis of
this spectral problem was accomplished in the
work~\cite{Gastine} where the experimental verification of the
calculated frequencies was conducted also by making use of radio
engineering measurements.

These studies enable one to separate all the dielectric sphere modes into the
{\it interior}  and {\it exterior} modes and, at the same time, into
the {\it volume} and {\it surface} modes. It is worth noting that all
the eigenfrequencies are complex
\begin{equation}
\label{3-15} \omega=\omega '-i\,\omega''\,{.}
\end{equation}
Thus we are dealing with "leaky modes". It is not surprising
because we are considering here  an open system~\cite{Open}  (a dielectric
ball and outer unbounded space).

The classification of the modes as the interior and exterior  ones
relies on the investigation of the behaviour of a given
eigenfrequency in the limit $\varepsilon \to \infty$. The modes are
called ''interior'' when the product $k\,a= \sqrt
{\varepsilon}\,\omega \,a/c$ remains finite in the limit
$\varepsilon \to \infty$, provided the imaginary part of the
frequency ($\omega ''$) tends to zero. The modes are referred to
as ''exterior'' when the product $k\,a/\sqrt{\varepsilon}=
\omega\,a/c$ remains finite with growing $\omega''$. In the first
case the frequency equations for a dielectric sphere (\ref{3-3}) and
(\ref{3-4}) tend to  (\ref{3-13}) and (\ref{3-14}) and in the
second case they tend to (\ref{2-7}) and  (\ref{2-8}). The
order of the root obtained will be denoted by the index $r$ for
interior modes and by $r'$ for exterior modes. Thus
${\rm TE}_{\,lr}$ and ${\rm TM}_{\,lr}$  denote the interior TE-
and TM-modes, respectively,  while ${\rm TE}_{\,lr'}$ and
${\rm TM}_{\,lr'}$ stand for the exterior TE- and TM-modes.

For fixed $l$ the number of the modes of exterior type is
limited because the frequency equations for exterior
oscillations of a perfectly conducting sphere (\ref{2-7}) and
(\ref{2-8}) have finite number of solutions~\cite{NFLS}. In view
of this, the number of exterior TE- and TM-modes is given by the
following rule. For even $l$ there are $l/2$ exterior TE-modes
and $l/2$  exterior TM-modes, for odd $l$ the number of the
modes ${\rm TE}_{\,l\,r'}$ is $(l+1)/2$ and the number of the
modes ${\rm TM}_{\,l\,r'}$ equals $(l-1)/2$.

An important parameter is the $Q$ factor
\begin{equation}
\label{3-16} Q_{{\rm rad}}=\frac{\omega'}{2\omega''}=
2\,\pi\,\frac{\mbox{stored energy}}{\mbox{radiated energy per
cycle}}\,{.}
\end{equation}
For exterior modes the value of $Q_{{\rm rad}}$ is always less than 1, hence
these modes can never be observed as sharp resonances. At the same time for
$\varepsilon $ greater than 5, the $Q_{{\rm rad}}$ for interior modes is
greater than 10 and it can reach very high values when
$\varepsilon \to \infty$.

In the problem at hand the losses due to the radiation can be
disregarded unlike the Ohmic losses. Indeed, the external source
of electromagnetic energy (cellular telephone) compensates the
radiation losses. While the Ohmic losses lead to the temperature
rising in human tissues.

For physical implications  more important is the classification
in terms of {\it volume} or {\it surface} modes according to
whether $r> l$ or $l> r$.  For volume modes the electromagnetic
energy is distributed in the whole volume of the sphere while in
the case of surface modes the energy is located close by the
sphere surface. The exterior modes are the first roots of the
characteristic equations and it can be shown that they are
always surface modes. \noindent
\begin{figure}[th]
\noindent \centerline{
\includegraphics[width=75mm]{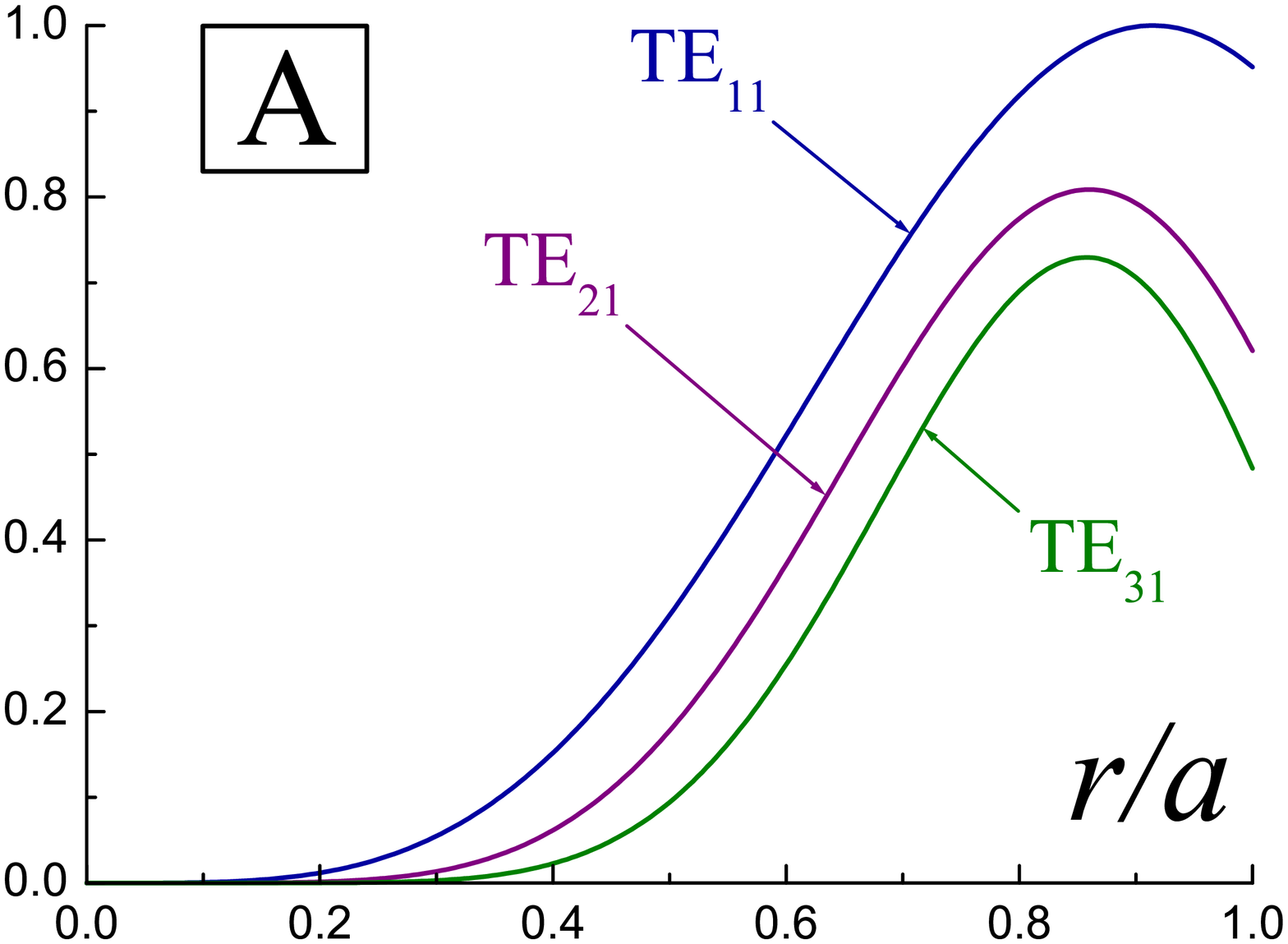}
\hspace{10mm}
\includegraphics[width=75mm]{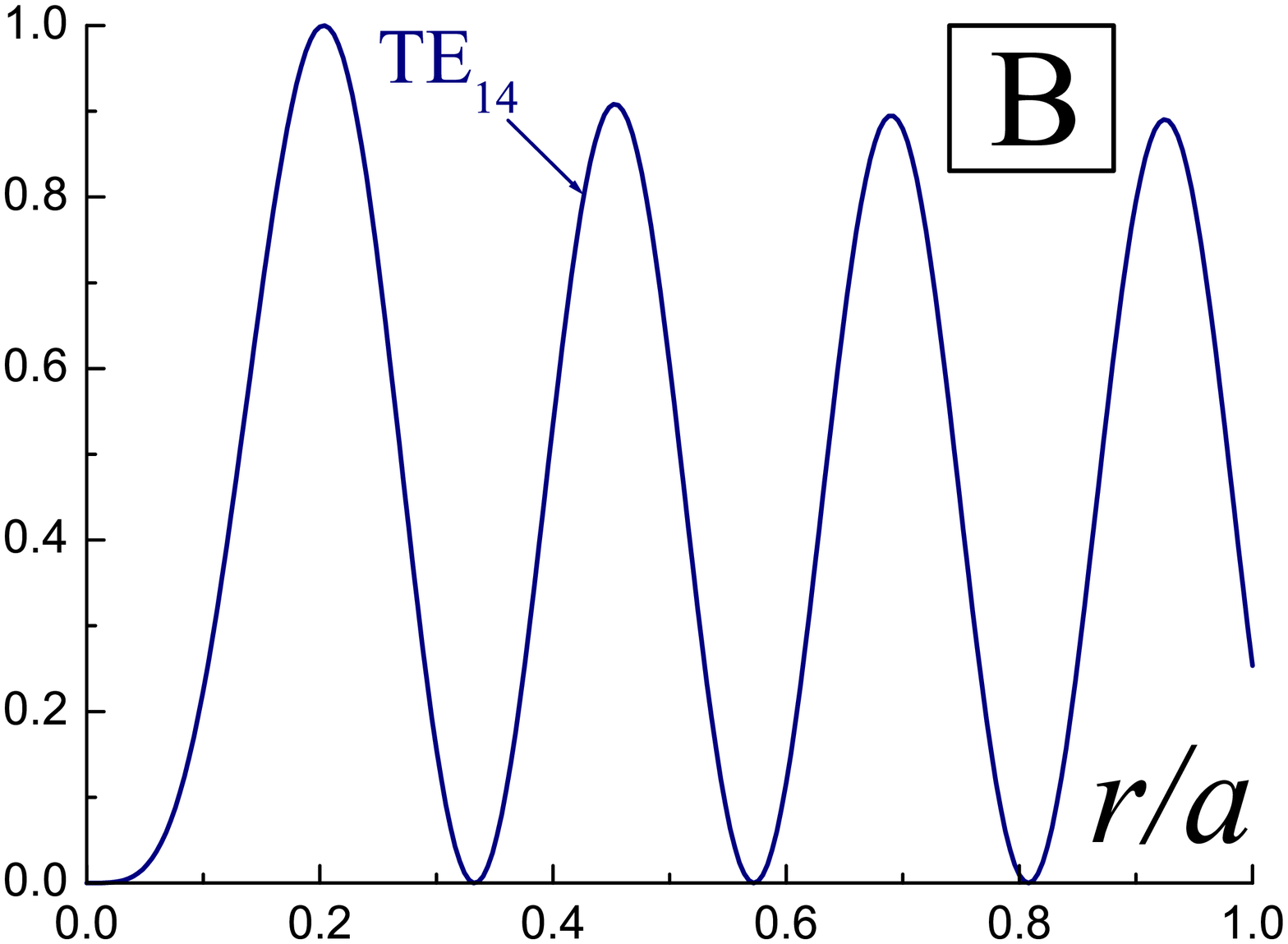}}
\caption{Electric energy density $r^2\,E_{\text{t}}^2$ for the
surface (A) and volume (B) TE-modes of a dielectric sphere with
$\varepsilon =40$ placed in vacuum.} \label{Plot:s-v-modes}
\end{figure}

Figure 1 shows a typical spatial behaviour of the surface and
volume modes of a dielectric sphere.

Thus a substantial part of the sphere modes (about one half) belong to the interior
surface modes. It is
important that respective frequencies are the {\it first} roots of
the characteristic equations.

In order to escape the confusion, it is worth noting here that the
surface modes in the problem in question obey the same boundary
conditions at the sphere surface  and when $r\to \infty$ as the
volume modes do. Hence,  these surface modes cannot be classified as the
evanescent surface waves propagating along the interface  between
two media (propagating waves along dielectric waveguides~\cite{Jackson}, surface
plasmon waves on the interface between metal bulk and adjacent
vacuum~\cite{Raether,BPN} and so on).  When describing the evanescent waves one imposes
the requirement of their exponential decaying away from interface
between two media. In this respect the evanescent surface wave
differ from the modes in the bulk.

{\bf Features of dielectric sphere spectrum and their
applications.}
%and estimation of the health hazard of portable telephones}
 The parts of human body (for example, head)
have the  eigenfrequencies of electromagnetic oscillations like
any compact body. In particular, one can anticipate that the
eigenfrequencies of human head are close to those of a
dielectric sphere with radius $a\approx 8$~cm and permittivity
$\varepsilon \approx 40$ (for human brain $\varepsilon =44.1$
for  900~MHz and $\varepsilon =42.8$ for  1.5~GHz \cite{WF}). By
making use of the results of calculations conducted in the work
\cite{Gastine} one can easily obtain the eigenfrequencies of a
dielectric sphere with the parameters  mentioned above. For
${\rm TE}_{l1}$ modes with $l=1,2,3$ we have, respectively, the
following frequencies: 280~MHz, 420~MHz, and 545~MHz. For ${\rm
TM}_{l1}$ modes with $l=1,2,3$ the resonance frequencies are
425~MHz, 540~MHz, and 665~MHz. The imaginary parts of these
eigenfrequencies are very small so the $Q$ factor in
(\ref{3-16}) responsible for radiation is greater than 100.

These eigenfrequencies belong to a new GSM 400 MHz frequency band
which is now being standardized by the European Telecommunications
Standards Institute.  This band was primarily used in Nordic
countries, Eastern Europe, and Russia in a first generation of
mobile phone system prior to the introduction of GSM.

    Due to the Ohmic losses the resonances of a dielectric sphere
 in question are in fact broad and overlapping.
Indeed, the electric conductance $\sigma$ of the human brain is
rather substantial. According to the data  presented in
\cite{WF} $\sigma\simeq 1.0$~S/m. The eigenfrequencies  of a
dielectric dissipative sphere with allowance for a finite
conductance $\sigma $ can be found in the following way. As
known \cite{LL} the effects of $\sigma $ on electromagnetic
processes in a media possessing a common real dielectric
constant $\varepsilon$ are described by a complex dielectric
constant $\varepsilon_{{\rm diss}}$ depending on frequency
\begin{equation}
\label{4-1}
\varepsilon_{{\rm diss}} =\varepsilon +i\frac{4\pi \sigma}{\omega}\,{.}
\end{equation}
The eigenfrequencies $\omega$,  calculated for a real $\varepsilon $,
are related to eigenfrequencies $\omega_{{\rm diss}}$ for $\varepsilon_{{\rm diss}}$
by the formula~\cite{LL}
\begin{equation}
\label{4-2} \omega_{{\rm diss}} =\left (
\frac{\varepsilon}{\varepsilon_{{\rm diss}} }\right
)^{1/2}\omega \simeq \omega -2\pi\, i
\,\frac{\sigma}{\varepsilon}\,{.}
\end{equation}
The corresponding factor $Q_{{\rm diss}}$ is
\begin{equation} \label{4-3} Q_{{\rm diss}}=\frac{\omega'_{{\rm diss}}}{2
\omega''_{{\rm diss}}}\simeq
\frac{\varepsilon \, \omega}{4\pi\, \sigma}\,{.}
\end{equation}
Substituting in this equation the values
$\omega /2\pi =0.5\cdot 10^9\;{\rm Hz}, \quad \varepsilon =40,
\quad \sigma = 1\,{\rm S/m}=9\cdot 10^9\; {\rm s}^{-1}$ one finds
\begin{equation}
\label{4-4}
Q_{{\rm diss}}\simeq\frac{20}{18}\simeq 1\,{.}
\end{equation}

In  view of such substantial Ohmic loses the resonance
enhancement of the oscillation amplitude inside  a human head
will not occur. However when the frequency of a mobile telephone
coincides with the eigenfrequency of the head the distribution
of electric and magnetic fields inside the head will be
described by the corresponding normal mode which may be a
surface mode or a volume one~\cite{Vaynstayn}.

Proceeding from this we propose the following test of numerical
calculations used when estimating the potential hazard of
cellular telephones. The test consists in simulation of the
temperature distribution corresponding to the surface and volume
modes in the framework of pertinent calculation schemes. For
simplicity, the test calculations could be accomplished for a
dielectric sphere (instead of a human head) with lower
conductivity  in comparison with that for a human brain (in
order to enhance the effect). The distributions of electric and
magnetic fields and the temperature distribution inside the
sphere should be calculated for two eigenfrequencies of the
sphere, namely,  one frequency corresponds to  surface mode and
another one belongs to volume mode. The distributions obtained
should  conform, at least qualitatively,  to the spatial
behaviour of respective electromagnetic normal modes (see
Fig.~1).

{\bf Conclusion.} Detailed analysis of electromagnetic spectra
of a dielectric sphere enables us to propose an independent
accuracy test of complicated numerical calculations conducted
when estimating the potential health hazard due to use of
cellular telephones. This test will permit one to make certain
of  a proper handling of the electromagnetic shape resonances of
a human head in these studies.

\begin{acknowledgments}
 This paper was completed during the visit of on of the authors
(VVN) to Salerno University. It is his pleasant duty to thank G.\
Scarpetta and G.\ Lambiase for the kind hospitality extended to
him. VVN was supported in part by the Russian Foundation for Basic
Research (Grant No.\ 06-01-00120). The financial support of INFN
is acknowledged. The authors are indebted to A.V.~Nesterenko for
preparing the figure.
\end{acknowledgments}


\begin{thebibliography}{99}
\bibitem{Health} International Commission on Non-Ionizing
Radiation Protection, IC-NIRP statement -- Health issues
related to the use of hand-held radiotelephones and base
transmitters,  Health Phys.  {\bf 70}, 587 (1996).
\bibitem{Sernelius} B.E.\ Sernelius,   Europhys.\ Lett.\ {\bf 60},  643 (2002).
\bibitem{WF}J.\ Wang, O.\ Fujiwara,  IEEE Trans.\ Microwave Theory and Techniques,
{\bf 47},   1528 (1999).
\bibitem{Whittaker} E.T.~Whittaker,   Proc.\ London Math.\ Soc. {\bf 1},  367 (1904).
\bibitem{Nisbet} A.~Nisbet,  Proc.\ Roy.\ Soc.\ London  A  {\bf 231},  250 (1955).
\bibitem{Stratton} J.A.\ Stratton,  {\i Electromagnetic Theory}  (McGraw-Hill, New York, 1941).
\bibitem{Jackson} J.D.\ Jackson, {\it Classical Electrodynamics}
3rd ed.  (Wiley, New York, 1999).
\bibitem{Sommerfeld} A.~Sommerfeld,   {\it Partial Differential Equations of
Physics} (Academic Press, New York, 1949).
\bibitem{NFLS} V.V.~Nesterenko, A.~Feoli, G.\ Lambiase, and G.\ Scarpetta, hep-th/0512340, v2.
%\eprint{gr-qc/9909058}.
\bibitem{Bar} V.V.\ Nesterenko, J.\ Phys.\ A: Math.\ Gen.  {\bf 39},  6609 (2006).
\bibitem{Nollert} H.-P. Nollert,  Class.\ Quantum Grav.\ {\bf 16}, R159 (1999).
\bibitem{FN} V.P.\ Frolov, I.D.\  Novikov,  {\it Black Hole Physics}  (Kluwer
Academic Publishers, Dordrecht, 1998).
\bibitem{KS} K.D.\ Kokkotas, B.G.\  Schmidt,   Living Rev.\ Relativity, {\bf 2},  1
 (1999);  gr-qc/9909058.
\bibitem{AS} M.~Abramowitz, I.~Stegun,  eds., {\it Handbook of Mathematical Functions}
(Dover, New York,  1972).
\bibitem{Debye} P.~Debye,  Ann.\ Phys.\ (Leipzig), {\bf 30},
57 (1909);  G.~Mie,   Ann.\ Phys.\ (Leipzig),  {\bf 25}, 377 (1908).
\bibitem{Gastine} M.~Gastine, L.~Courtois, J.L.~Dormann,
IEEE Trans.\ Microwave Theory and Techniques,
{\bf 15}, 694 (1967).
\bibitem{Open} K.-H.~Li, Phys.\ Reports, {\bf 134}, 1 (1986).
\bibitem{Raether} H.~Raether, {\it Surface Plasmons} (Springer, Berlin,  1988).
\bibitem{BPN}  M.~Bordag, I.G.~Pirozhenko, V.V.\  Nesterenko,  J.\ Phys.\ A:
 Math.\ Gen. {\bf 38},  11027 (2005).
\bibitem{LL} L.L.\ Landau, E.M.\ Lifshitz, {\it Electrodynamics of Continuouis Media,
Course of Theoretical Physics}  Vol.~ 8  (Pergamon, New York,  1960).
\bibitem{Vaynstayn} L.A.\ Vaynstayn, {\it Theory of open resonators and open waveguides}
(Sovetskoe Radio, Moscow,  1966, English translation 1969).
\end{thebibliography}
\end{document}